%
\let\includefigures=\iftrue
%
%
%
\newfam\black
\input harvmac.tex
\input rotate
\input epsf
\input xyv2
\noblackbox
\includefigures
\message{If you do not have epsf.tex (to include figures),}
\message{change the option at the top of the tex file.}
\def\figin{\epsfcheck\figin}\def\figins{\epsfcheck\figins}
\def\epsfcheck{\ifx\epsfbox\UnDeFiNeD
\message{(NO epsf.tex, FIGURES WILL BE IGNORED)}
\gdef\figin##1{\vskip2in}\gdef\figins##1{\hskip.5in}
\else\message{(FIGURES WILL BE INCLUDED)}%
\gdef\figin##1{##1}\gdef\figins##1{##1}\fi}
\def\DefWarn#1{}

\def\figinsert{\goodbreak\midinsert}
\def\ifig#1#2#3{\DefWarn#1\xdef#1{fig.~\the\figno}
\writedef{#1\leftbracket fig.\noexpand~\the\figno}%
\figinsert\figin{\centerline{#3}}\medskip\centerline{\vbox{\baselineskip12pt
\advance\hsize by -1truein\noindent\footnotefont{\bf Fig.~\the\figno:} #2}}
\bigskip\endinsert\global\advance\figno by1}
\else
\def\ifig#1#2#3{\xdef#1{fig.~\the\figno}
\writedef{#1\leftbracket fig.\noexpand~\the\figno}%
\global\advance\figno by1}
\fi

\def\subsubsec#1{\bigskip\noindent{\it #1}}
\def\yboxit#1#2{\vbox{\hrule height #1 \hbox{\vrule width #1
\vbox{#2}\vrule width #1 }\hrule height #1 }}
\def\fillbox#1{\hbox to #1{\vbox to #1{\vfil}\hfil}}
\def\ybox{{\lower 1.3pt \yboxit{0.4pt}{\fillbox{8pt}}\hskip-0.2pt}}

\def\rightarrowbox#1#2{
  \setbox1=\hbox{\kern#1{${ #2}$}\kern#1}
  \,\vbox{\offinterlineskip\hbox to\wd1{\hfil\copy1\hfil}
    \kern 3pt\hbox to\wd1{\rightarrowfill}}}

\def\p{\partial}

\def\Tr{{{\rm Tr~ }}}

\def\ad{{\rm ad~ }}

\def\bigvev#1{\left\langle{#1}\right\rangle}

\def\II{\relax{I\kern-.10em I}}

\def\bar{\overline}

\def\IZ{\relax\ifmmode\mathchoice
{\hbox{\cmss Z\kern-.4em Z}}{\hbox{\cmss Z\kern-.4em Z}}
{\lower.9pt\hbox{\cmsss Z\kern-.4em Z}}
{\lower1.2pt\hbox{\cmsss Z\kern-.4em Z}}\else{\cmss Z\kern-.4em
Z}\fi}
\def\IB{\relax{\rm I\kern-.18em B}}
\def\IC{{\relax\hbox{$\inbar\kern-.3em{\rm C}$}}}
\def\ID{\relax{\rm I\kern-.18em D}}
\def\IE{\relax{\rm I\kern-.18em E}}
\def\IF{\relax{\rm I\kern-.18em F}}
\def\IG{\relax\hbox{$\inbar\kern-.3em{\rm G}$}}
\def\IGa{\relax\hbox{${\rm I}\kern-.18em\Gamma$}}
\def\IH{\relax{\rm I\kern-.18em H}}
\def\II{\relax{\rm I\kern-.18em I}}
\def\IK{\relax{\rm I\kern-.18em K}}
\def\IN{\relax{\rm I\kern-.18em N}}
\def\IP{\relax{\rm I\kern-.18em P}}

%
\def\inbar{\,\vrule height1.5ex width.4pt depth0pt}

\def\p{\partial}

\font\cmss=cmss10 \font\cmsss=cmss10 at 7pt
\def\IR{\relax{\rm I\kern-.18em R}}

\def\lp10{l_P^{10}}
\def\lp11{l_P^{11}}
\def\R11{R_{11}}

\baselineskip 14pt
%
%
%

\lref\KonishiOne{ K.~Konishi, ``Anomalous Supersymmetry
Transformation Of Some Composite Operators In Sqcd,'' Phys.\
Lett.\ B {\bf 135}, 439 (1984).
}

\lref\KonishiTwo{ K.~i.~Konishi and K.~i.~Shizuya, ``Functional
Integral Approach To Chiral Anomalies In Supersymmetric Gauge
Theories,'' Nuovo Cim.\ A {\bf 90}, 111 (1985).
}

\lref\DijkgraafTwo{ R.~Dijkgraaf and C.~Vafa, ``Matrix models,
topological strings, and supersymmetric gauge theories,''
arXiv:hep-th/0206255.
}

\lref\DijkgraafThree{ R.~Dijkgraaf and C.~Vafa, ``On geometry and
matrix models,'' arXiv:hep-th/0207106.
}

\lref\DijkgraafOne{ R.~Dijkgraaf and C.~Vafa, ``A perturbative
window into non-perturbative physics,'' arXiv:hep-th/0208048.
}

\lref\DijkgraafZan{ R.~Dijkgraaf, M.~T.~Grisaru, C.~S.~Lam,
C.~Vafa and D.~Zanon, ``Perturbative Computation of Glueball
Superpotentials,'' arXiv:hep-th/0211017.
}

\lref\WittenOne{F.~Cachazo, M.~R.~Douglas, N.~Seiberg and E.~Witten, ``Chiral Rings and Anomalies in Supersymmetric Gauge Theories,'' JHEP {\bf 0212}, 071 (2002) [arXiv:hep-th/0211170].}

\lref\WittenTwo{F.~Cachazo, N.~Seiberg and E.~Witten, ``Phases of {\cal N}=1 
Supersymmetric Gauge Theories and Matrices,'' JHEP {\bf 0302}, 042 (2003) [arXiv:hep-th/0301006].}

\lref\WittenThree{E.~Witten, ``Chiral Ring of Sp(N) and SO(N) Supersymmetric 
Gauge Theory in Four Dimensions'', [arXiv:hep-th/0302194].}

\lref\WittenFour{F.~Cachazo, N.~Seiberg and E.~Witten, ``Chiral Ring and Phases of Supersymmetric Gauge Theories,'' [arXiv:hep-th/0303207].}

\lref\Seiberg{N.~Seiberg, ``Adding fundamental matter to 'Chiral Rings and Anomalies in Supersymmetric Gauge Theories','' JHEP {\bf 0301}, 061 (2003) [arXiv:hep-th/0212225].}

\lref\Veneziano{G.~Veneziano and S.~Yankielowicz, ``An Effective Lagrangian For The Pure {\cal N}=1 Supersymmetric Yang-Mills Theory,'' Phys.\ Lett.\ B {\bf 113}, 231 (1982)}

\lref\Parvizi{R.~Abbaspur, A.~Imaanpur and S.~Parvizi, ``{\cal N}=2 $SO(N)$ SYM Theory from Matrix Model,'' [arXiv:hep-th/0302083].}

\lref\Obers{R.~A.~Janik and N.~A.~Obers, ``$SO(N)$ Superpotential, Seiberg-Witten Curves and Loop Equations,'' [arXiv:hep-th/0212069].}

\lref\Corrado{S.~K.~Ashok, R.~Corrado, N.~Halmagyi, K.~D.~Kennaway and C.~Romelsberger, ``Unoriented Strings, Loop Equations, and {\cal N}=1 Superpotentials from Matrix Models,'' [arXiv:hep-th/0211291].}

\lref\Ahn{C.~Ahn and S.~Nam, ``{\cal N}=2 Supersymmetric $SO(N)$/$Sp(N)$ Gauge Theories from Matrix Model,'' [arXiv:hep-th/0301203], 
C.~h.~Ahn and S.~Nam,
 ``Supersymmetric SO(N) gauge theory and matrix model,''
[arXiv:hep-th/0212231], 
C.~h.~Ahn and Y.~Ookouchi,
``Phases of N = 1 supersymmetric SO / Sp gauge theories via matrix model,''
JHEP {\bf 0303}, 010 (2003)
[arXiv:hep-th/0302150],  
C.~h.~Ahn,
``Supersymmetric SO(N)/Sp(N) gauge theory from matrix model: Exact  mesonic vacua,''
[arXiv:hep-th/0301011].
}

\lref\Feng{B.~Feng, ``Geometric Dual and Matrix Theory for $SO(N)$/$Sp(N)$ Gauge Theories,'' [arXiv:hep-th/0212010].}

\lref\OzOne{H.~Ita, H.~Nieder and Y.~Oz, ``Perturbative Computation of Glueball Superpotentials for $SO(N)$ and $USp(N)$,'' [arXiv:hep-th/0211261].}

\lref\Ookouchi{Y.~Ookouchi and Y.~Watabiki, ``Effective Superpotentials for $SO$/$Sp$ with Flavor from Matrix Models,'' [arXiv:hep-th/0301226].}

\lref\Kraus{P.~Kraus and M.~Shigemori, ``On the Matter of the Dijkgraaf-Vafa Conjecture,'' [arXiv:hep-th/0303104].}

\lref\OzTwo{A.~Brandhuber, H.~Ita, H.~Nieder, Y.~Oz and C.~Romelsberger, ``Chiral Rings, Superpotentials and the Vacuum Structure of {\cal N}=1 Supersymmetric Gauge Theories,'' [arXiv:hep-th/0303001].}

\lref\FujiWD{
H.~Fuji and Y.~Ookouchi,
``Comments on effective superpotentials via matrix models,''
JHEP {\bf 0212}, 067 (2002)
[arXiv:hep-th/0210148].
}

\lref\all{
R.~Boels, J.~de Boer, R.~Duivenvoorden and J.~Wijnhout,
``Nonperturbative superpotentials and compactification to three  dimensions,''
[arXiv:hep-th/0304061], 
S.~G.~Naculich, H.~J.~Schnitzer and N.~Wyllard,
``Cubic curves from matrix models and generalized Konishi anomalies,''
[arXiv:hep-th/0303268], 
A.~Klemm, K.~Landsteiner, C.~I.~Lazaroiu and I.~Runkel,
``Constructing gauge theory geometries from matrix models,''
[arXiv:hep-th/0303032], 
M.~Matone,
``Seiberg-Witten duality in Dijkgraaf-Vafa theory,''
Nucl.\ Phys.\ B {\bf 656}, 78 (2003)
[arXiv:hep-th/0212253], 
D.~Berenstein,
``Quantum moduli spaces from matrix models,''
Phys.\ Lett.\ B {\bf 552}, 255 (2003)
[arXiv:hep-th/0210183], 
N.~Dorey, T.~J.~Hollowood, S.~Prem Kumar and A.~Sinkovics,
``Exact superpotentials from matrix models,''
JHEP {\bf 0211}, 039 (2002)
[arXiv:hep-th/0209089]. 
A.~Gorsky,
``Konishi anomaly and N = 1 effective superpotentials from matrix models,''
Phys.\ Lett.\ B {\bf 554}, 185 (2003)
[arXiv:hep-th/0210281], 
V.~Balasubramanian, B.~Feng, M.~x.~Huang and A.~Naqvi,
``Phases of N = 1 supersymmetric gauge theories with flavors,''
[arXiv:hep-th/0303065].
}

\lref\SeibergJQ{
N.~Seiberg,
``Adding fundamental matter to 'Chiral rings and anomalies in  supersymmetric gauge theory',''
JHEP {\bf 0301}, 061 (2003)
[arXiv:hep-th/0212225].
}

\lref\KrausTwo{
P.~L.~Cho and P.~Kraus,
``Symplectic SUSY gauge theories with antisymmetric matter,''
Phys.\ Rev.\ D {\bf 54}, 7640 (1996)
[arXiv:hep-th/9607200], 
C.~Csaki, W.~Skiba and M.~Schmaltz,
``Exact results and duality for Sp(2N) SUSY gauge theories with an  antisymmetric tensor,''
Nucl.\ Phys.\ B {\bf 487}, 128 (1997)
[arXiv:hep-th/9607210].
}

%
%
\newbox\tmpbox\setbox\tmpbox\hbox{\abstractfont RUNHETC-2002-09}
\Title{\vbox{\baselineskip12pt\hbox to\wd\tmpbox{\hss
hep-th/0304119}
}}
{\vbox{\centerline{Effective Superpotentials via Konishi Anomaly}}}
\smallskip
\centerline{L.~F.~Alday,$^{1,2,3}$ M.~Cirafici$^{1,2}$}
\smallskip
\bigskip
\centerline{$^1$ {\it SISSA, Trieste, Italy}}
\medskip
\centerline{$^2$ {\it INFN, Sez. di Trieste, Trieste, Italy}}
\medskip
\centerline{$^3$ {\it The Abdus Salam ICTP, Trieste, Italy}}
\bigskip
\vskip 1cm
 \noindent
We use Ward identities derived from the generalized Konishi anomaly in order 
to compute effective superpotentials for $SU(N)$, $SO(N)$ and $Sp(N)$ 
supersymmetric gauge theories coupled to matter in various representations. In
 particular we focus on cubic and quartic tree level superpotentials. With 
this technique higher order corrections to the perturbative part of the 
effective superpotential can be easily evaluated.

\Date{April 2002}
%

%
%
\newsec{Introduction}

The nonperturbative dynamics of supersymmetric gauge theories is a rich and 
fascinating subject. These theories exhibit various phenomena, such as gaugino
 condensation and confinement, a complete understanding of whose is still 
lacking. Recently a promising step toward this direction has been made by 
Dijkgraaf and Vafa \refs{\DijkgraafOne,\DijkgraafTwo,\DijkgraafThree} who 
conjectured that the effective superpotentials for a wide class of ${\cal N}=1$
theories can be computed by summing planar diagrams in a related matrix model. 
Arguments supporting this conjecture were given using chiral superspace 
techniques in \DijkgraafZan\ and using anomalies in \WittenOne. In particular 
this last result led the authors of \WittenOne\ to further interesting 
developments \refs{\WittenTwo,\WittenFour,\SeibergJQ,\WittenThree}. 
Many aspects of this conjecture have been widely studied in the 
last months \all.

In \WittenOne, a generalization of the Konishi anomaly \refs{\KonishiOne,
\KonishiTwo} has been considered, leading to a set of identities with the same structure of the loop equations of a matrix model. Using 
these identities the authors were able to derive the effective superpotentials
 of \DijkgraafOne\ up to an integration constant (basically the 
Veneziano-Yankielowicz superpotential \Veneziano).  The aim of this work is 
to follow their approach to compute effective superpotentials in some specific 
cases 
for $SU(N)$, $SO(N)$ and $Sp(N)$\foot{Here we use conventions such that $N$ is an even number, i.e. the rank of the group is ${N \over 2}$.} with matter in various representations.

Some superpotentials for $SO(N)$ and $Sp(N)$ have already been computed in the
 framework of \refs{\DijkgraafOne,\DijkgraafZan}, see for example \refs{\Parvizi
\Obers\Corrado\Ahn\Feng\OzOne\Ookouchi-\Kraus}. For adjoint matter, the results
 obtained reflect the charge of the orientifold plane used in the geometric 
engineering of the gauge theory. For discussions on $SU(N)$ see for example 
\OzTwo.

This paper is organized as follows: in the second section we review some results
 from \WittenOne\ and in section 3 we compute the generalized Konishi anomaly 
for other groups, namely $SO(N)$, $Sp(N)$ and $SU(N)$ and write down the equations (analog to the loop equations of the matrix model) we will use to compute effective superpotentials. In the 
fourth section we perform explicit computations of effective superpotentials 
for cubic and quartic interactions with matter in various representations. Finally in the last section we discuss our results and propose some further 
developments.

\newsec{A Fairy Tale of an Anomaly}

In this section we will briefly summarize some results of \WittenOne. 

Let us begin with an $U(N)$ theory with matter $\Phi$ in the adjoint representation and a tree level superpotential
\eqn\potential{W(\Phi)=\sum_{k=0}^n {g_k\over k+1}\Tr \Phi^{k+1}}
as in \WittenOne. This theory has a natural ring structure, the chiral ring, defined by the equivalence classes of gauge invariant chiral operators modulo 
$\{\bar Q_{\dot \alpha},\dots\}$. It can be proven that it is generated by operators of the form $\Tr\,\Phi^k$,
$\Tr\,\Phi^k W_\alpha$, and $\Tr\, \Phi^kW_\alpha W^\alpha$, where $W^\alpha$ is the gauge superfield. An important element of the chiral ring is the glueball
superfield $S=-{1\over 32 \pi^2} \Tr\, W_\alpha W^\alpha$, which is believed to describe the low energy dynamics of the theory.
 Given a tree level superpotential, the effective superpotential as a function of the glueball superfield $S$ can be found, restricting ourselves to the chiral ring, by means of Ward identities following from a generalization of the Konishi anomaly.   

In \WittenOne\ the most general variation of $\Phi$ in the chiral ring was considered
\eqn\variation{\Phi=f(\Phi, W_\alpha)}
where $f$ is a general holomorphic function, leading to the generalized Konishi anomaly\foot{Here and in the following we will consider $W(\Phi)$ as a matrix every time it appears inside a trace.}
\eqn\anomaly{
\bar D^2 J_f = \Tr f(\Phi,W_\alpha) {\p W(\Phi)\over \p \Phi} + 
 \sum_{ijkl} A_{ij,kl} {\p f(\Phi,W_\alpha)_{ji} \over \p \Phi_{kl}}}
where\foot{With $\ad V$ we mean the adjoint representation: 
 $(\ad V~ \Phi)^i{}_j = V^i{}_k \Phi^k{}_j - \Phi^i{}_k V^k{}_j$. }
$$J_f = \Tr \bar\Phi e^{\ad V} f(\Phi,W_\alpha)$$
and
\eqn\formA{A_{ij,kl}= {1\over 32\pi^2} \left[W_\alpha,
 \left[W^\alpha, T_{lk}\right]\right]_{ij}}
 $T_{lk}$ being the generators of the gauge group ($U(N)$ in this case). Note that at this point this is quite general and a change in the gauge group will reflect only in the explicit form of the generators $T_{lk}$. For $U(N)$ we have
$(T_{lk})_{ij} = (e_{lk})_{ij} = \delta_{il} \delta_{jk}$ and
\eqn\genkonishian{
\bar D^2 J_f = \Tr f(\Phi,W_\alpha) {\p W(\Phi)\over \p \Phi} +
 {1\over 32\pi^2}
 \sum_{i,j} \left[W_\alpha, \left[W^\alpha, {\p f\over\p \Phi_{ij}}\right]
 \right]_{ji} .
}
Finally, taking the vacuum expectation value, we find
\eqn\wardId{
\bigvev{ \Tr f(\Phi,W_\alpha) {\p W\over\p\Phi} } =
 - {1\over 32\pi^2}
\bigvev{ \sum_{i,j} \left( \left[W_\alpha,\left[W^\alpha,
         {\p f(\Phi,W_\alpha) \over \p \Phi_{ij}}
           \right]\right]\right)_{ji} }.
}
Now, we define
\eqn\defgen{\eqalign{ T(z) &= \sum_{k\ge 0} z^{-1-k} \Tr
\Phi^k = \Tr {1\over z-\Phi}  \cr 
R(z) &= -{1\over 32\pi^2}\Tr
{W_\alpha W^\alpha}{1\over z-\Phi} }}
Taking
\eqn\vargen{
\delta \Phi_{ij} = f_{ij}(\Phi,W_{\alpha }) = R(z)_{ij}
}
using the Konishi anomaly and the algebraic relation
\eqn\algeid{ \sum_{i,j}
\left[\chi_1,\left[\chi_2, {\p\over\p \Phi_{ij}}
 {\chi_1 \chi_2\over z-\Phi}\right]\right]_{ij} =
 \left(\Tr {\chi_1 \chi_2\over z-\Phi} \right)^2
}
which holds if $\chi_1^2=\chi_2^2=0$ and $\left[ \Phi , \chi_\alpha \right]=0$,
 one can obtain
\eqn\Requation{
R^2(z) = W'(z) R(z) + {1\over 4}f(z) 
}
$f(z)$ being a polynomial of degree $n-1$ in $z$.
Solving this equation, we obtain
\eqn\Rsoln{
2 R(z) =  W'(z) - \sqrt{W'(z)^2 + f(z)}}
Analogously, by taking
\eqn\vargenT{
\delta \Phi_{ij} = f_{ij}(\Phi) = T(z)_{ij}
}
one finds
\eqn\Tequation{ 2 R(z) T(z) = W'(z) T(z) + {1\over 4}c(z)}
where $c(z)$, like $f(z)$, is a polynomial of degree $n-1$. This equation can 
be used together with \Rsoln\ in order to derive a closed equation for $T(z)$
\eqn\Teqn{
T(z)=-{1\over4}{c(z)\over \sqrt{W'(z)^2 + f(z)}}}
As explained in \WittenOne, expanding \Teqn\ in powers of $1\over z$ we can extract the vacuum expectation values of the operators $ \Tr \Phi^k$ that can be integrated to obtain the effective superpotential up to a constant of integration independent of the couplings (but in general dependent on $S$), using the relation
\eqn\DweffDg{{\partial W_{eff} \over \partial g_k}= \bigvev{ \Tr {\Phi^{k+1}\over k+1}}}
Then, the general strategy is to write \genkonishian\ for the gauge group under
 consideration, obtain the generalization of \Requation\ and \Tequation, 
solve for $T(z)$ as in \Teqn\ and finally extract the superpotential 
(as explained before).

The information about the chosen vacuum is encoded in the explicit form of the 
 denominator of equation \Teqn. In general, the choice of the function $f(z)$ in the curve
$$ y^2 = W'(z)^2 + f(z) $$
determines how the gauge group is broken and selects the vacuum. In particular, for the case of unbroken gauge group, only one of the $n$ zeros of $W'(z)$ 
splits in a branch cut (around the vacuum), while the others will only get shifted. 
This means that the curve factorizes as $$  y^2 = W'(z)^2 + f(z)= Q(z)^2 
\left( z + \alpha + \beta \right) \left( z + \alpha - \beta \right) $$ where
 the $n-1$ unsplitted zeros of $y$ are contained in the polynomial $Q(z)$.


\newsec{The Konishi anomaly for other groups }


In this section, we will derive the Konishi anomaly and the equation for $T(z)$ for $SO(N)$ (in some detail) and $Sp(N)$ with matter in the adjoint and 
symmetric (antisymmetric for $Sp(N)$), both traceful and traceless, 
representations and finally for $SU(N)$ with matter in the adjoint representation.

Let us begin with the case of an $SO(N)$ gauge theory with adjoint matter and 
evaluate explicitly \anomaly. We take the generators of $SO(N)$ to be 
$ T_{lk} = \left(e_{lk}-e_{kl} \right)$ with $(e_{lk})_{ij}=\delta_{il} 
\delta_{jk}$. First of all, we note that the identity 
\algeid\ holds due to the spinorial properties 
of $\chi_ \alpha$ and is independent of the generators up to numerical factors.
 As can be easily 
checked the equation for $R(z)$ \Requation\ then becomes 
$$ {1 \over 2} R^2(z) = W'(z) R(z) + {1\over 4}f(z) $$
whose solution is
\eqn\Req{2 R(z) =  2W'(z) - 2 \sqrt{W'(z)^2 + {f(z) \over 2}}}
Now let us focus on the equation for $T(z)$ \Tequation\ and restrict 
ourselves to variations of the
 form\foot{Properly speaking one should add also the term $1 \over z+ \Phi$ since $\delta \Phi$ has to be an element of $SO(N)$ in the adjoint representation, that is to say an antisymmetric matrix. However it can be checked that it  will contribute exactly as the previous, giving only an overall factor of $2$. Because of this it will be omitted in the following analysis.}
\eqn\varT{
\delta \Phi = f(\Phi)=-{1 \over 32 \pi^2} {1 \over z- \Phi}
}
Then the equation for the anomaly gives
\eqn\anomalySO{\eqalign{
\bar D^2 J_f = &-{1 \over 32 \pi^2} \Tr {1 \over z- \Phi} {\p W(\Phi)\over \p \Phi} \cr &+ {1\over 32\pi^2} {1 \over 4}
 \sum_{ijkl} \left[W_\alpha,
 \left[W^\alpha,  \left(e_{lk} - e_{kl} \right) \right]
 \right]_{ij} \left( {1 \over z- \Phi} 
  \left (e_{kl} - e_{lk} \right) 
{1 \over z- \Phi} \right)_{ji}}}
Let us focus on the second term on the right hand side
\eqn\rhsSO{\eqalign{&{1 \over 32 \pi^2} {1 \over 4}
 \sum_{ijkl} \left[W_\alpha,
 \left[W^\alpha,  \left( e_{lk} - e_{kl} \right) \right] 
\right]_{ij} \left( {1 \over z- \Phi} 
  \left( e_{kl} - e_{lk} \right) {1 \over z- \Phi} 
\right)_{ji} = \cr &{1 \over 32\pi^2} {1 \over 4} 
\left(4  \Tr {W_\alpha W^\alpha \over z - \Phi} \Tr {1 \over z - \Phi}
-8 \Tr \left( W_\alpha W^\alpha {1 \over z - \Phi} \left( {1 \over z - \Phi} \right)^T \right)
-4  \Tr {W_\alpha \over z - \Phi} \Tr {W^\alpha \over z - \Phi} \right)
 }}
where we have used the commutation properties of the operators in the chiral 
ring. Now, being $\Phi$ an antisymmetric matrix, we have
\eqn\transS{\left( {1 \over z - \Phi} \right)^T = {1 \over z + \Phi} }
Next we use the identity
$$
 {1 \over z - \Phi} {1 \over z + \Phi} = {1 \over 2z} \left({1 \over z - \Phi}
 + {1 \over z + \Phi} \right)
$$
in order to write
$$
 \Tr W_\alpha W^\alpha {1 \over z - \Phi} \left( {1 \over z - \Phi} \right)^T 
= {1 \over z} \Tr W_\alpha W^\alpha {1 \over z - \Phi}
$$
Taking expectation values of \anomalySO\ and using the definitions \defgen\
we have
$$
W'(z)T(z)+{1 \over 4}c(z) =  R(z)T(z) - 2 {R(z) \over z}  
$$
Using the relation \Req\ we finally obtain the equation for $T(z)$
\eqn\TeqSO{T(z)=-{1\over4}{c(z)\over \sqrt{W'(z)^2 + f(z)}}-{2 \over z} 
 { W'(z)- \sqrt{W'(z)^2 + f(z)} \over \sqrt{W'(z)^2 + f(z)}}
}
Here we absorbed a factor of ${1 \over 2}$ in a redefinition of $f(z)$ ( 
we will always use this convention when speaking about $SO(N)$ and $Sp(N)$). 
As previously explained, from \TeqSO\ we can obtain the effective superpotential for an $SO(N)$ gauge theory with adjoint matter and tree level superpotential \potential.

Now let us consider the same gauge theory but with matter in the symmetric representation (that is, $\Phi$ is now a symmetric matrix and we use a symmetric representation for the $SO(N)$ basis).
In this case \transS\ becomes
$$
\left( {1 \over z - \Phi} \right)^T = {1 \over z - \Phi}
$$
and
\eqn\derzmphi{\eqalign{ \Tr W_\alpha W^\alpha {1 \over z - \Phi} \left( {1 \over z - \Phi} \right)^T 
&= \Tr W_\alpha W^\alpha \left( {1 \over z - \Phi} \right) ^2 \cr &= - { d \over dz} \left(
\Tr W_\alpha W^\alpha {1 \over z - \Phi} \right) }}
Again, from \genkonishian, \defgen\ and using now \derzmphi\ one finds
\eqn\loopTtrSO{W'(z)T(z)+{1 \over 4}c(z) =  R(z)T(z) - 2 R'(z) } 
and the equation for $T(z)$ becomes
\eqn\TeqSosymm{T(z)=-{1\over4}{c(z)\over \sqrt{W'(z)^2 + f(z)}} - 2  { {
d \over dz} \left( W'(z)- \sqrt{W'(z)^2 + f(z)} \right) \over \sqrt{W'(z)^2 + f(z)}}
}
To complete our discussion about $SO(N)$, let us consider now $\Phi$ in the 
traceless symmetric representation. All we have to do is to take the previous 
results and subtract the trace of $\Phi$. For instance, \varT\ will now become
\eqn\notrace{
\delta \Phi = f(\Phi)=-{1 \over 32 \pi^2} \left( {1 \over z- \Phi} -
{1 \over N} \Tr {1 \over z- \Phi} \right)
}
This will not produce any change in \rhsSO\ (since the trace part is proportional to the identity matrix and it is entering in the commutator); the only modifications will arise in the left hand side of \wardId\ which now becomes
$$
-{1 \over 32 \pi^2} \Tr \left( {1 \over z- \Phi} - {1 \over N} \Tr \left( {1 \over z- \Phi} \right) \right) {\p W(\Phi)\over \p \Phi}
$$
and in the equation for $R(z)$ \Requation\ which now reads\foot{Remember that we are taking vacuum expectation values; properly speaking $W'(\Phi)$ 
has to be understood as $\langle W'(\Phi) \rangle$.}
\eqn\RtrlessSO{R^2(z) = \left(W'(z) - {1 \over N}  W'(\Phi) \right)
 R(z) + {1\over 4}f(z)}
Now equation \loopTtrSO\ becomes
$$
T(z) \left( W'(z)- {1 \over N}  W'(\Phi) \right) + {1 \over 4}c(z) = 
 R(z)T(z) - 2 R'(z)
$$
Finally we can write the equation for $T(z)$ for matter in the symmetric 
traceless representation
\eqn\TeqSoTrless{\eqalign{T(z) = &-{1\over4}{c(z)\over \sqrt{ \left( W'(z)- {1 \over N}  W'(\Phi) \right)^2 + f(z)}} \cr &-2  { {
d \over dz} \left( \left( W'(z)- {1 \over N}  W'(\Phi) \right) - 
\sqrt{ \left( W'(z)- {1 \over N}  W'(\Phi) \right)^2 + f(z)} \right) \over 
\sqrt{ \left( W'(z)- {1 \over N}  W'(\Phi) \right)^2 + f(z)}}
}}

Now we will focus on an $Sp(N)$ gauge theory with matter in the adjoint 
(symmetric) and in the antisymmetric (both traceful and traceless) 
representations. With symmetric (antisymmetric) we 
mean that $\Phi$ has to be considered as a matrix $MJ$ where $M$ is a 
symmetric (antisymmetric) matrix and $J$ is the invariant antisymmetric 
tensor of $Sp(N)$. We take the generators of $Sp(N)$ as $ 
\left(e_{lk}+e_{kl} \right)$ with $(e_{lk})_{ij}=\delta_{il} \delta_{jk}$. The 
analysis for the $Sp(N)$ case is almost identical to the one for the $SO(N)$ 
case, the only change being the sign in the generators (and of course the 
different properties of the matrices representing the field $\Phi$, since the 
antisymmetric invariant $J$ will enter in the intermediate steps). Because of
 this we will only state our results.
For matter in the symmetric representation the equation for $T(z)$ becomes
\eqn\TeqSp{T(z)=-{1\over4}{c(z)\over \sqrt{W'(z)^2 + f(z)}} + {2 \over z} 
 { W'(z)- \sqrt{W'(z)^2 + f(z)} \over \sqrt{W'(z)^2 + f(z)}}
}
for matter in the antisymmetric traceful representation
\eqn\TeqSpanti{T(z)=-{1\over4}{c(z)\over \sqrt{W'(z)^2 + f(z)}} + 2  { {
d \over dz} \left( W'(z)- \sqrt{W'(z)^2 + f(z)} \right) \over \sqrt{W'(z)^2 + f(z)}}
}
and finally for matter in the antisymmetric traceless representation
\eqn\TeqSpTrless{\eqalign{T(z) = &-{1\over4}{c(z)\over \sqrt{ \left( W'(z)- {1 \over N}  W'(\Phi) \right)^2 + f(z)}} \cr &+2  { {
d \over dz} \left( \left( W'(z)- {1 \over N}  W'(\Phi) \right) - 
\sqrt{ \left( W'(z)- {1 \over N}  W'(\Phi) \right)^2 + f(z)} \right) \over 
\sqrt{ \left( W'(z)- {1 \over N}  W'(\Phi) \right)^2 + f(z)}}
}}

As a last example, let us consider the $SU(N)$ gauge group with matter in the adjoint representation. This is basically equivalent to consider an $U(N)$ 
gauge theory subtracting the trace as in \notrace\ (remember that the term 
containing the trace will not produce any modification when entering in a 
commutator). Then, one can easily find
\eqn\TeqSU{
T(z) = -{1\over4}{c(z)\over \sqrt{ \left( W'(z)- {1 \over N}  W'(\Phi) 
\right)^2 + f(z)}}
}


\newsec{The effective superpotential}


In this section we apply the previous results in order to find the effective 
superpotential for $SO(N)$, $Sp(N)$ and $SU(N)$ gauge theories with quartic 
and cubic superpotential and matter in various representations.

As already mentioned, the general strategy is to write down the equation for $T(z)$ for every particular case, expand it in powers of ${1 \over z}$ and extract the vacuum expectation values of the operators $ \langle\Tr{\Phi^k} \rangle$, from where the effective superpotential can be obtained by using equation 
\DweffDg.

\subsec{Quartic superpotential}

\subsubsec{Sp(N)/SO(N) with matter in the antisymmetric/symmetric representation}

\

Let us suppose the following tree level superpotential:

\eqn\quarticpot{ 
W(\Phi)={m \over 2} \Tr \Phi^2+{g \over 4} \Tr \Phi^4
}

As seen in the previous section we obtain the following equation for $T(z)$

\eqn\Tquartic{
T(z)=-{1 \over 4}{c(z) \over {\sqrt{W'(z)^2+f(z)}}} + 2 \epsilon 
{{W''(z)-(\sqrt{W'(z)^2+f(z)})'}\over{\sqrt{W'(z)^2+f(z)}}} 
}
Where $ \epsilon = \pm 1 $ for $Sp(N)/SO(N)$ and $c(z)$ and $f(z)$ are polynomials of degree 2
$$f(z)=\sum_{i=0}^{2} f_i z^i$$ and $$c(z)=\sum_{i=0}^{2} c_i z^i$$ 

The denominator of both terms in equation \Tquartic\ can be factorized as explained previously. We impose (see for example \Ookouchi):

\eqn\factorization{
W'(z)^2+f(z)=g^2 (z^2 - k^2)^2 (z^2 - 4\mu^2) 
}
From this condition we arrive to the following expressions:

\eqn\conditions{\eqalign{
k& = \sqrt { - {{m+2g \mu^2} \over g}} \cr
\mu& = \sqrt {-{gm + \sqrt { g^2 \left( -3 f_2+m^2 \right) } \over 6g^2}}
}}

Imposing condition \factorization\ gives a system of equations for $k$ and 
$\mu$ with a set of solutions. Note that we choose the particular $k$ and 
$\mu$ tending to $ \sqrt{- {m \over g}} $ and $0$ for $f_2$ going to zero 
(that is the classical limit). This means we place the branch cut around zero.
In order to have the correct asymptotic behavior of $R(z)$ for large $z$  
( $R(z) \sim {S \over z}$ , see eq. \defgen) it 
can be shown that\foot{The difference with \WittenOne\ is due to our 
redefinition of $f(z)$.}  $f_2=-2 g S$. Similarly the correct asymptotic 
behavior of $T(z)$ for large $z$ ( $T(z) \sim {N \over z}$ )  sets 
$c_2=-4 g N$. $c_0$ and $c_1$ can be 
found by asking the condition that $T(z)$ has no poles in $k$ and $-k$ 
(i.e. we choose our vacuum around $\Phi=0$ and the gauge group remains 
unbroken) or, equivalently:

\eqn\noresidue{\eqalign{ {1\over 2\pi
i}\oint_{C_k} dz \,T(z)=0 \cr
{1\over 2\pi
i}\oint_{C_{-k}} dz \,T(z)=0
}}
For the present case we obtain:

\eqn\cquartic{\eqalign{ 
c_0 &= 4 \left( 2 \epsilon \left( m + gk \left( 3k -2 \sqrt{k^2 
-4 \mu^2 } \right) \right) +gk^2N \right) \cr 
c_1 &= 0
}} 
Next, we expand $T(z)$ in powers of ${1 \over z}$ and obtain:

\eqn\weffqeq{
{{\partial W_{eff}}\over {\partial m}}= {2 \epsilon \over g} \left( m+3g 
\mu^2 + \sqrt{\left(m+6g \mu^2 \right) \left( m+2g\mu^2 \right) } \right) 
 + \mu^2 N 
 }
This expression can be expanded in powers of $S$ and integrated in order to obtain the effective superpotential up to any given order; for instance up to 
fourth order, it reads
\eqn\weffq{\eqalign{ 
W_{eff} =& {1 \over 2}  \left(- 2\,\epsilon +N \right) S  \log m +{g \over 8m^2} \left( - 10\,\epsilon +3N \right) S^2 -
 {g^2 \over 16m^4} \left( - 38\,\epsilon + 9N \right) S^3 
\cr &+ {g^3 \over 96m^6} \left( - 662\,\epsilon + 135 N \right)S^4 + \dots
}}

Several comments are in order. Having obtained $W_{eff}$ integrating with respect to $m$, the result is correct up to a function of $g$ and $S$ (a part of which is the Veneziano-Yankielowicz superpotential); we could have chosen 
the coefficient of the term ${1 \over z^4}$ and integrated 
with respect to $g$. As the perturbative part of the potential depends only on
 the ratio ${g \over m^2}$ by dimensional reasons \WittenOne, a function of 
only one of the coupling constants cannot contribute. For the same reason, from now on, we will only consider the ${1 \over z^3}$ term.  

\subsubsec{Sp(N)/SO(N) with matter in the adjoint representation}

\

This case is completely analogous to the case studied before.
 Now the equation for $T(z)$ reads:
$$
T(z)=-{1\over4}{c(z)\over \sqrt{W'(z)^2 + f(z)}}+ \epsilon {2 \over z} 
 { W'(z)- \sqrt{W'(z)^2 + f(z)} \over \sqrt{W'(z)^2 + f(z)}}
$$
Where again $c(z)$ and $f(z)$ are polynomials of degree 2. The denominator of \
both terms can be factorized as before (since it is the same) and we obtain 
the same values for the parameters $k$ and $ \mu $. Again $f_2=-2 g S$ 
and $c_2=-4 g N$ and the conditions \noresidue\ must be imposed. The values 
obtained for $c_0$ and $c_1$ are:

\eqn\cquartadj{\eqalign{ 
c_0 &= 4 \left( 2 \epsilon (gk^2+m) + g k^2 N \right) \cr
c_1 &= 0
}} 
Again, expanding $T(z)$ in powers of ${1 \over z}$ and and extracting the coefficient of ${1 \over z^3}$ we obtain:
\eqn\weffqadjm{
{{\partial W_{eff}}\over {\partial m}}= \mu^2 (N + 2 \epsilon )
}
that can be expanded in powers of $S$ and integrated with respect to $m$, 
to give:
\eqn\weffadjexp
{
W_{eff}=(N+2 \epsilon) \left( {S \over 2 } \log m + {{3 g S^2}\over{8 m^2}}-{{9 g^2 S^3}\over{16 m^4}}+{{45 g^3 S^4}\over{32 m^6}} - {{567 g^4 S^5} \over {128 m^8}} + { {5103 g^5 S^6} \over {320 m^{10}}}  + \dots \right)
}

This result agrees with the one of \FujiWD, where the effective superpotentials
 were evaluated using both matrix model techinques and in terms of closed 
strings on Calabi-Yau geometry with fluxes. 

\subsec{Cubic superpotential}

\subsubsec{SO/Sp with traceful symmetric/antisymmetric matter}

\

Now the superpotential under consideration takes the form:
\eqn\cubicpot{ 
W(\Phi)={m \over 2} \Tr \Phi^2 + {g \over 3} \Tr \Phi^3
}
The equation for $T(z)$ reads exactly as in \Tquartic\ but now $c(z)$ and 
$f(z)$ are polynomials of degree $1$. Again we factorize the denominator of 
both terms, now as follows:

\eqn\factorization{
W'(z)^2+f(z) = g^2 (z-k)^2 (z+a+b) (z+a-b)
}
In this case, the parameters $a$, $b$ and $k$ are complicated functions of $m$, $g$ and $f_1$; because of this we will only write their expansion in powers
of $S$:
\eqn\abkexp{\eqalign{
k & = - {m \over g} + a \cr
a & = {g \over m^2} S + 3 {g^3 \over m^5} S^2 + 16 {g^5 \over m^8} S^3 +
 105 {g^7 \over m^{11}} S^4 + 768 {g^9 \over m^{14}} S^5 + 6006 {g^{11} \over 
m^{17}} S^6+ 49152 {g^{13} \over m^{20}} S^7 + \dots \cr
b & = \sqrt {S \over 2m} \left( 2 + 2 {S g^2 \over m^3}  + 9 { S^2 g^4 
\over m^6}
 + 55 { S^3 g^6 \over m^9} + {1547 \over 4} { S^4 g^8 \over m^{12}}   
 + {11799 \over 4} { S^5 g^{10} \over m^{15}} + {189805 \over 8} 
{ S^6 g^{12} \over m^{18}} + \dots \right) 
}}

Note that in the classical limit (that is $S \rightarrow 0$ ) the parameters $a$ and $b$ go to zero, while $k$ tends to its classical value $ - {m \over g}$. 
Again the asymptotic behavior of $R(z)$ and $T(z)$ imposes $f_1=-2g S$ and $c_1=-4 g N$, and, as before, $c_0$ is set by the condition that $T(z)$ does not have a pole at $z=k$:

\eqn\noresiduek{
{
1\over 2\pi
i}\oint_{C_k} dz \,T(z)=0 
}
and from this
$$ c_0 = 8 \epsilon \left( 2gk + g \sqrt{\left(a - b +k \right) 
\left(a + b +k \right) } +m \right) 
 + 4 g k N  $$
As before, $T(z)$ can be expanded in powers of ${1 \over z}$ and we can integrate the coefficient of ${1 \over z^3}$ with respect to $m$ in order to obtain the effective superpotential.

 We stress that without too much difficult one can obtain the result up to the desired order. For instance, up to seventh order:

\eqn\Weffcubicadj{\eqalign{
W_{eff}= &{1 \over 2}  \left(-2\,\epsilon + N \right)S\log m -
{g^2 \over 2m^3} \left(- 3\,\epsilon + N \right)S^2 -{1 \over 12} 
{g^4 \over m^6} \left(- 59\,\epsilon + 16N \right)S^3 \cr &- {1 \over 24} 
{g^6 \over m^9} \left(- 591\,\epsilon + 140N \right)S^4 - {1 \over 16} 
{g^8 \over m^{12}} \left( -{4775 \over 2}\,\epsilon + 512N \right)S^5 
\cr &- {1 \over 80} {g^{10} \over m^{15}} \left( - 80763\,\epsilon + 16016N 
\right) S^6  - {1 \over 96} {g^{12} \over m^{18}} \left( - 704809\,\epsilon 
+ 131072N \right) S^7 + \dots
}}
 
Note that our results are in perfect agreement, up to $S^5$, with the ones of \Kraus\ found  using the matrix model perturbative approach of \DijkgraafZan .

\subsubsec{SO/Sp with traceless symmetric/antisymmetric matter}

\

In the case of matter in the traceless representation, the equation of $T(z)$ reads:
\eqn\TeqSoTrless{\eqalign{T(z) = &-{1\over4}{c(z)\over \sqrt{ \left( W'(z)- {1 \over N}  W'(\Phi) \right)^2 + f(z)}} \cr & + 2\,\epsilon  { {
d \over dz} \left( \left( W'(z)- {1 \over N}  W'(\Phi) \right) - 
\sqrt{ \left( W'(z)- {1 \over N}  W'(\Phi) \right)^2 + f(z)} \right) \over 
\sqrt{ \left( W'(z)- {1 \over N}  W'(\Phi) \right)^2 + f(z)}}
}}
Here, as we will see, the strategy we follow is different, due to the fact that we are considering the traceless representation and that $\Tr(\Phi^2)$ appears explicitly in the denominator of $T(z)$. First we factorize the denominator in the usual way:

\eqn\tracelessden{
(W'(z)-{1 \over N}W'(\Phi))^2+f(z) = g^2 (z-k)^2 (z^2+a z + b)
}
As we are in the traceless representation we have:

\eqn\traceless{\eqalign{
\Tr \Phi &= 0 \cr
W'(\Phi) &= g \Tr \Phi^2
}}

The polynomial $c(z)$ can be fixed as before, and again $f_1=-2g S$. The condition of $\Tr \Phi = 0$ implies that the coefficient of $1 \over z^2$ in the expansion of $T(z)$ must be zero. We can use this condition together with the conditions of factorization in order to obtain a system of equations from where $\Tr \Phi^2$ can be evaluated. Equivalently the traceless condition can be used to
determine $c_0$ and equation \noresiduek\ together with the conditions from 
factorization can be used to determine $\Tr \Phi^2$.
As before we obtain from this the effective superpotential. It should be stressed that such evaluation can be done at any desired number of loops, without many technical complications. For the effective superpotential one finds:

\eqn\wefftraceless{\eqalign{ 
W_{eff}= & 
  {(N - 2\epsilon){S \over 2} \log m}+
{{ g^2 (-\epsilon\,N+4) S^2} \over { 2 N m^3}}
 \cr & +{{ g^4 (160\,\epsilon -24 N - N^2\,\epsilon) S^3} \over {12 m^6 N^2}} 
+{{ g^6 (3584 - 256\,\epsilon N  -36 N^2 - \epsilon\,N^3)S^4} \over 
{24 m^9 N^3}} \cr &+ {g^8  \left( 67584\,\epsilon - 
704 N^2\,\epsilon - 48 N^3 - N^4\,\epsilon \right) S^5  \over {32m^{12} N^4}}
 \cr &+  {7 g^{10} \left( 1171456 +  79872 \,\epsilon N - 8320 N^2 - 1280 
\epsilon N^3 -60 N^4 -\epsilon N^5 \right) S^6 \over 240 m^{15} N^5} + \dots
}}
Note that these results agree with the ones of \Kraus; however using 
this method is easier to compute higher loop corrections.

\subsubsec{ $SU(N)$ with adjoint matter}

\

Recently in \Kraus\ it was found that for a cubic potential, like \cubicpot, 
the perturbative part of the effective superpotential is zero up to terms 
of order $S^4$, due to cancellations in the diagrammatic evaluation. In this 
paragraph we will show that the generalized Konishi anomaly implies that
the perturbative part of $W_{eff}$ is exactly vanishing to all orders.
Let us consider equation \TeqSU\ and expand it in powers of ${1 \over z}$
\eqn\sunexp{\eqalign{
T(z) & = -{1\over4}{c_0 + c_1 z  \over \sqrt{ \left( W'(z)- {1 \over N}  
W'(\Phi) \right)^2 + f(z)}} \cr & =
-{c_1 \over 4g }{1 \over z} + {1 \over 4g} \left( - c_0 + {c_1 m \over g}
\right) {1 \over z^2}  
+ {\cal O} \left( {1 \over z^3} \right)   
\cr & = {N \over z} + { \langle \Tr \Phi \rangle \over z^2 } + 
{\cal O} \left( {1 \over z^3} \right) 
}}

From the terms of order ${1 \over z}$ we find the familiar condition 
$c_1 = -4 gN$. Considering the term ${1 \over z^2}$ and imposing the 
tracelessness of $\Phi$ in $SU(N)$ we obtain the relation 
$c_0 = {c_1 m \over g}$.
Again the denominator of \sunexp\ can be factorized as in \tracelessden
$$ (W'(z)-{1 \over N}W'(\Phi))^2+f(z) = g^2 (z-k)^2 (z^2+a z + b) $$
then the condition \noresiduek\ gives the following relation
$$ a=0 $$
With this condition only odd powers of ${1 \over z}$ will be present in the 
expansion of $T(z)$; in particular
$$ {\p W_{eff} \over \p g}=0 $$
from which we see that the perturbative part of the effective superpotential 
is identically zero (remember that the perturbative part depends only on a 
specific ratio of $m$ and $g$, in this case $ {g^2 \over m^3} $). 

We stress that the vanishing of the superpotential is a particular 
characteristic of the cubic superpotential. One can easily check that for a 
quartic tree level superpotential, a non zero result is obtained. 


\newsec{Conclusions}


In this paper we used the generalized Konishi anomaly approach of \WittenOne\ 
in order to compute some effective superpotentials for the gauge groups $SO(N)$, $Sp(N)$ and $SU(N)$ with quartic and cubic tree level superpotential and 
matter in various representations. Our results are summarized in the appendix.

Our results agree with previous literature, where superpotentials were computed
 summing planar diagrams in the matrix model context. The method used in this
paper does not rely on any diagrammatic expansion, but simply on Ward identities that allow us to write closed expressions for the generating functions of
correlators; because of this higher corrections are easily evaluated by
power expansions without having to draw any diagram. 

In particular, the cancellation of the perturbative part of $W_{eff}$ for $SU(N)$ gauge group and cubic superpotential, that in \Kraus\ was checked diagrammatically up to fourth order in $S$, can be easily shown to hold at any order.

In \Kraus, it was found within the matrix model approach, for the $Sp(N)$ 
gauge theory, a result disagreeing with previous gauge theory computations 
\KrausTwo\ at $h={N \over 2} + 1$ loops, $h$ being the dual Coxeter number. 
This led the authors of \Kraus\ to suggest that the general Dijkgraaf-Vafa 
 conjecture needs to be modified, at least in this particular case. 
In this 
paper we reproduced their matrix model prediction from the gauge theory side, 
within the formalism of \WittenOne. 
Then, unfortunatly, we are not able to give any hint on this intriguing 
problem. 

Possible further developments are to apply these techniques to more general superpotentials or other gauge groups (maybe exceptional); in some cases it may be even possible to find the {\it exact} effective superpotential. 

\centerline{\bf Acknowledgements}

We are particularly grateful to E.~Gava and K.~S.~Narain for useful discussions and comments. We also thank J.~David, R.~Dijkgraaf, C.~Maccaferri, D.~Mamone,
 and L.~Mazzucato for discussions.

\appendix{A}{Summary of results}

In this appendix we collect our result, stating first the equation for $T(z)$ 
and then the corresponding superpotential for each case. Remember that 
$\epsilon= \pm 1$ for $Sp(N)$/$SO(N)$.

\subsec{$SU(N)$ with adjoint matter and cubic interactions}

$$
T(z) = -{1\over4}{c(z)\over \sqrt{ \left( W'(z)- {1 \over N}  W'(\Phi) 
\right)^2 + f(z)}}
$$

$$ {\p W_{eff} \over \p g}=0 $$

\subsec{$Sp(N)$/$SO(N)$ with matter in the antisymmetric/symmetric traceful 
representation}

$$
T(z)=-{1 \over 4}{c(z) \over {\sqrt{W'(z)^2+f(z)}}} + 2 \epsilon 
{{W''(z)-(\sqrt{W'(z)^2+f(z)})'}\over{\sqrt{W'(z)^2+f(z)}}}
$$
For a cubic superpotential
\eqn\appSOone{\eqalign{
W_{eff}= &{1 \over 2}  \left(-2\,\epsilon + N \right)S\log m -
{g^2 \over 2m^3} \left(- 3\,\epsilon + N \right)S^2 -{1 \over 12} 
{g^4 \over m^6} \left(- 59\,\epsilon + 16N \right)S^3 \cr &- {1 \over 24} 
{g^6 \over m^9} \left(- 591\,\epsilon + 140N \right)S^4 - {1 \over 16} 
{g^8 \over m^{12}} \left( -{4775 \over 2}\,\epsilon + 512N \right)S^5 
\cr &- {1 \over 80} {g^{10} \over m^{15}} \left( - 80763\,\epsilon + 16016N 
\right) S^6  - {1 \over 96} {g^{12} \over m^{18}} \left( - 704809\,\epsilon + 
131072N \right) S^7 + \dots
}}
For a quartic superpotential
\eqn\appSOtwo{\eqalign{ 
W_{eff} =& {1 \over 2}  \left(- 2\,\epsilon +N \right) S  \log m +{g \over 8m^2} \left( - 10\,\epsilon +3N \right) S^2 -
 {g^2 \over 16m^4} \left( - 38\,\epsilon + 9N \right) S^3 
\cr &+ {g^3 \over 96m^6} \left( - 662\,\epsilon + 135 N \right)S^4 + \dots
}}

\subsec{$Sp(N)$/$SO(N)$ with matter in the antisymmetric/symmetric traceless 
representation and cubic interactions}

\eqn\appSOthree{\eqalign{T(z) = &-{1\over4}{c(z)\over \sqrt{ \left( W'(z)- {1 \over N}  W'(\Phi) \right)^2 + f(z)}} \cr & + 2\,\epsilon  { {
d \over dz} \left( \left( W'(z)- {1 \over N}  W'(\Phi) \right) - 
\sqrt{ \left( W'(z)- {1 \over N}  W'(\Phi) \right)^2 + f(z)} \right) \over 
\sqrt{ \left( W'(z)- {1 \over N}  W'(\Phi) \right)^2 + f(z)}}
}}

\eqn\appSOfive{\eqalign{ 
W_{eff}= & 
  {(N - 2\epsilon){S \over 2} \log m}+
{{ g^2 (-\epsilon\,N+4) S^2} \over { 2 N m^3}}
 \cr & +{{ g^4 (160\,\epsilon -24 N - N^2\,\epsilon) S^3} \over {12 m^6 N^2}} 
+{{ g^6 (3584 - 256\,\epsilon N  -36 N^2 - \epsilon\,N^3)S^4} \over 
{24 m^9 N^3}} \cr &+ {g^8  \left( 67584\,\epsilon - 
704 N^2\,\epsilon - 48 N^3 - N^4\,\epsilon \right) S^5  \over {32m^{12} N^4}}
 \cr &+  {7 g^{10} \left( 1171456 +  79872 \,\epsilon N - 8320 N^2 - 1280 
\epsilon N^3 -60 N^4 -\epsilon N^5 \right) S^6 \over 240 m^{15} N^5} + \dots
}}

\subsec{$Sp(N)$/$SO(N)$ with matter in the adjoint representation and quartic interactions}

$$
T(z)=-{1\over4}{c(z)\over \sqrt{W'(z)^2 + f(z)}}+ \epsilon {2 \over z} 
 { W'(z)- \sqrt{W'(z)^2 + f(z)} \over \sqrt{W'(z)^2 + f(z)}}
$$

\eqn\appSOfour
{
W_{eff}=(N+2 \epsilon) \left( {S \over 2 } \log m + {{3 g S^2}\over{8 m^2}}-{{9 g^2 S^3}\over{16 m^4}}+{{45 g^3 S^4}\over{32 m^6}} - {{567 g^4 S^5} \over {128 m^8}} 
+ { {5103 g^5 S^6} \over {320 m^{10}}} + \dots \right)
}

\listrefs

\end